\documentclass[a4paper,11pt]{article}
\pdfoutput=1 

\usepackage{jcappub} 

\usepackage[T1]{fontenc} 

\title{\boldmath  Spherically symmetric static solutions, newtonian potential and degrees of freedom of a Nonlocal action}

 \author{Utkarsh Kumar}
 \author{and Sukanta Panda}
 \affiliation{Indian Institute of Science Education and Research, Bhopal,\\ Bhauri (462066), India}

\emailAdd{utkarshk@iiserb.ac.in}
\emailAdd{sukanta@iiserb.ac.in}

\abstract{Nonlocal terms in the Einstein Hilbert(EH) action appears as IR corrections in effective theory of quantum gravity. Here we have considered such an action keeping the terms which are quadratic in Ricci Scalar. We obtain the solution for a spherical symmetric. We conduct an analysis for linearised gravity action and find the Newtonian potentials of linearized gravity action and count the degrees of freedom. Finally we reproduce the results for few known examples found in literature.}

\begin{document}
\maketitle
\flushbottom

\section{Introduction}
There are overwhelming evidences from the supernovae observations \cite{Riess:1998cb} of accelerated expansion of universe \cite{Rubin:2016iqe}. To explain the cosmic acceleration, 
one adds the cosmological constant(cc) in the local Einstein Hilbert(EH) action. However, the presence of cc in the action suffers from the serious issue of fine-tuning. In order to avoid such issues we take an alternative approach where our action contains nonlocal terms in addition to EH action. These nonlocal modified gravity models can be employed to study the cosmology in both IR and UV regimes. In order to study the cosmology in IR regime, first  phenomenogical nonlocal model was introduced by Deser and Woodard \cite{Deser:2007jk,Woodard:2014iga} by invoking a simple term of form $ R f(\frac{R}{\Box})$ in 
classical EH action of GR. After that many nonlocal models were introduced for a better understanding 
of accelerated solutions in these theories \cite{Koivisto:2008dh,Dodelson:2013sma,Dirian:2014xoa,Nersisyan:2016hjh,
Cusin:2015rex,Tsamis:2009ja,Tsamis:2010ph,Romania:2012av,Codello:2016neo,Codello:2016xhm,
Belgacem:2017cqo,Park:2017zls,Amendola:2017qge,Kumar:2018chy,Kumar:2018pkb}.

Origin of these nonlocal corrections can be understood  by treating quantum gravity as an effective field theory \cite{Codello:2016elq,Codello:2015pga,Maggiore:2016fbn,Donoghue:1993eb,Donoghue:1994dn}. These nonlocal terms arises when one performs the heat kernel expansion of effective action. In this work, we study the static spherically symmetric solution of a particular nonlocal model. This particular nonlocal model with action is \cite{Codello:2016neo} 
\begin{equation}
S = \int d^{4}x \sqrt{-g}M_{p}^{2}[ R + aR^{2} + cR\frac{1}{\Box^{2}}R ],   \label{unified}
\end{equation}

where $a$ and $c$ are arbitrary parameters. Action (\ref{unified}) consists of dynamical 
variables $ R^{2},$ which plays significant role at early time evolution of the universe. It is argued in Ref. \cite{Starobinsky:1980te,Mukhanov:1981xt} that  addition of $ R^{2}$ term in EH action drives the inflation of our universe without adding any extra scalar field. The other term in action (\ref{unified})  plays the role of a cosmological constant. It is reported in ref\cite{Codello:2016neo} that evoloution of field equations corresponding to action (\ref{unified}) gives complete discription of universe at both early and late times. However it's worth mentioning that action (\ref{unified}) does not contain all the quadratic terms. The complete action up to $ 2^{nd} $ order in Ricci scalar obtained in a EFT of gravity is \cite{Codello:2016xhm, Codello:2015pga}
\begin{equation}
S = \int d^{4}x \sqrt{-g} M_{p}^{2}\Big[ R + a R^{2} + R \mathcal{F}\Big(\frac{-\Box}{m^{2}}\Big)R  + \chi C^{2} - C_{\mu\nu\rho\sigma} \mathcal{G}\Big( \frac{-\Box}{m^2}\Big) C^{\mu\nu\rho\sigma} - 2\Lambda \Big], 
\label{complteaction}
\end{equation}

where $\chi $ is another arbitrary parameter which may be fixed by data. $ \mathcal{F}$ and $\mathcal{G}$,  known as structure functions, are non-local in the low $ k $ regime. Expansion of $ \mathcal{F}$ in the limit $ m^{2} \gg -\Box $ takes the following form \cite{Codello:2016xhm}
\begin{equation}
\mathcal{F}\Big( -\frac{\Box}{m^{2}}\Big) = -b \frac{m^{2}}{-\Box} -\gamma \frac{m^{2}}{-\Box}\log \frac{-\Box}{m^{2}} - \delta  \log \frac{-\Box}{m^{2}} -\zeta \Big( \frac{m^{2}}{-\Box}\Big)^{2}.
\end{equation} 
Structure for $ \mathcal{G}$ is given in equations (77) to (79) of Ref. \cite{Codello:2015pga,Codello:2015mba}.

In our work, we will consider only the terms containing R while neglecting logarithmic term. Hence our action reduces to 
\begin{equation}
S = \int d^{4}x \sqrt{-g} \Big[  R + a R^2 + b R \frac{1}{\Box} R - c R \frac{1}{\Box^{2}}R \Big].  \label{intrestaction}
\end{equation}

Paper is organised as follows : In Section (\ref{sec2}) we set up all required equations to find solutions for a most general maximally static symmetric metric. Numerical solutions of these equations are obtained in Section (\ref{sec3}). Section (\ref{sec4}) is devoted to study the weak field implications of our action. Section (\ref{sec5}) and (\ref{sec6}) deal with calculation of newtonian potentials and counting the physical degree of freedom of the theory respectively. These calculations for some known examples are provided here. In last Section (\ref{sec8}) we summarize our findings. 

\section{Equivalent scalar-tensor action} \label{sec2}

We consider the following nonlocal action (\ref{intrestaction}) with matter part
\begin{equation}
S = \int d^{4}x \sqrt{-g} \Big[  R + a R^2 + b R \frac{1}{\Box} R - c R \frac{1}{\Box^{2}}R  +  2 \kappa^2\mathcal{L}_{m} \Big].
\end{equation}
Here  $a , b , c$  are the constants of  mass dimension $ -2,0,2$ respectively. $\mathcal{L}_{m} $ is matter part of action.

We introduce the Lagrange multipliers  P and Q in the action to rename the $ \Box^{-1} $  and $ \Box^{-2}$ in terms of auxillary scalar fields U and S. Here $U$ and $S$ satisfy

\begin{equation}
\Box^{-1}R = -U \hspace{0.3em}\Rightarrow \Box U = -R
\end{equation}  
and 
\begin{equation}
\Box^{-1}S = -U \hspace{0.3em} \Rightarrow \Box S = -U \hspace{0.3em} \Rightarrow S = \Box^{-2}R.
\end{equation}

Using above defnitions we rewrite our action as
\begin{equation}
S = \int d^{4}x \sqrt{-g} \Bigg[ R + a R^2 + b \Big\{ -UR + P( \Box U + R ) \Big\} + c \Big\{ -RS + P( \Box U + R ) + Q( \Box S + U ) \Big\} \Bigg], 
\label{action}
\end{equation}

Further by integrating by parts action (\ref{action}) becomes
\begin{equation}
S = \int d^{4}x \sqrt{-g} \Bigg [ R\Big( 1 + a R -\beta(U - P) -c( S - P)\Big) -(b + c)\nabla_{\sigma}P \nabla^{\sigma}U -c\Big(\nabla_{\sigma}Q \nabla^{\sigma}S  + QU\Big) \Bigg]   \label{raction}
\end{equation}

Varying  action (\ref{raction}) with respect to scalar fields U , S , P and Q respectively, we get

\begin{eqnarray}
(b + c)\Box P &=& b R -c Q    \label{peom},\\
\Box Q&=&R,    \label{qeom}   \\
\Box U &=& -R,    \label{ueom} \\
\Box S &=& -U.   \label{seom}
\end{eqnarray}

From equations (\ref{qeom}) and (\ref{ueom}) it turns out that $ U = -Q $ and equation \ref{peom} becomes
\begin{equation}
(b + c)\Box P = b R + c U    \label{peom1}.
\end{equation}

\hspace{2em} The final action (\ref{raction}) becomes
\begin{equation}
 S = \int d^{x} \sqrt{-g} \Bigg [ R\Big( 1 + a R -\beta(U - P) -c( S - P)\Big) -(b + c)\nabla_{\sigma}P \nabla^{\sigma}U - c\Big(\nabla_{\sigma}U \nabla^{\sigma}S  + U^2\Big) \Bigg].   \label{r1action}
 \end{equation}

 Varying action (\ref{r1action}) with respect to metric $ g_{\mu\nu} $, we get 
 \begin{equation}
 \begin{split}
 G_{\mu\nu}  & + 2a \Bigg\{ R\Big( G_{\mu\nu} + \frac{1}{4} R g_{\mu\nu} \Big) + \Big(g_{\mu\nu} \Box R -\nabla_{\mu}\nabla_{\nu}R \Big) \Bigg\} + \\& \Big( R_{\mu\nu} -\nabla_{\mu}\nabla_{\nu} + g_{\mu\nu} \Box -\frac{1}{2}g_{\mu\nu} R \Big)\Bigg\{ b\Big( P-U \Big) + c\Big( P-S \Big)\Bigg\} \\& + \frac{1}{2}g_{\mu\nu}g^{\sigma \lambda} \Bigg\{ \Big( (b -c)\partial_{\sigma} U \partial_{\lambda} P - c \partial_{\sigma}U \partial_{\lambda} S \Big) \Bigg\} + (b -c)\partial_{\mu} U \partial_{\nu} P + c \partial_{\mu} U \partial_{\nu} S  + \frac{1}{2} c g_{\mu\nu} U^2 = \kappa^2 T_{\mu\nu}. \label{GMUNU}
 \end{split}
 \end{equation}
 By taking trace of equation (\ref{GMUNU}) we obtain the constraint equation as 
 \begin{equation}
 R \Bigg\{ b \Big( P- U \Big) + c \Big(P-S \Big) + 1 \Bigg\} = 6 a \Box R + \Big( b - c \Big) \partial_{\sigma} U \partial_{\sigma} P  - c \partial_{\sigma} U \partial^{\sigma} S   + 2c U^2 . \label{traceqn}
 \end{equation}
 \subsection{Basic equations}
 Now we study the dynamics of these equations in static spherically symmetric space. We consider most general static spherically symmetric metric(sss) of the form
 \begin{equation}
 ds^2 = -e^{2\alpha(r)}dt^2 + e^{2\beta(r)}dr^2 + r^2(d\theta^2 + sin^2\theta d\phi^2)  \label{metric}
\end{equation} 
For this metric, different components of nonzero ricci tensor $R_{\mu\nu}$ and ricci scalar $R$ are 
\begin{eqnarray} 
R_{tt}& = & \frac{1}{r}\Bigg[ e^{2(\alpha-\beta)} \Big\{ r\alpha'^{2} + \alpha'\Big(2 - r\beta' \Big) + r\alpha'' \Big\}\Bigg],  \label{Rtt} \\
R_{rr}& =& -\Bigg[ \alpha'' + \alpha'^{2} - \frac{2\beta'}{r} -\alpha'\beta' \Bigg],  \label{Rrr}\\
R_{\theta\theta} & = & -e^{-2\beta}\Bigg[ 1 - e^{2\beta} +r\alpha' -r\beta' \Bigg], \label{Rth}  \\
R_{\phi\phi}& = & e^{-2\beta}sin^{2}\theta \Bigg[ -1 + e^{2\beta} -r\alpha' + r\beta' \Bigg],   \label{Rph} \\
R &= &\frac{2}{r^2} - e^{-2\beta}\Bigg[ \frac{2}{r^2} + \frac{4}{r}\big(\alpha' -\beta'\Big) + 2\Big(\alpha'' + \alpha'^2 -\alpha'\beta' \Big)\Bigg] .  \label{R} 
\end{eqnarray}

Specialization of invariant d'Alembertaion operator for the mertic equation (\ref{metric}) gives
\begin{equation}
\Box f = \Big[ f'' + \Big( \alpha'-\beta' + \frac{2}{r}\Big) f'\Big] \label{Box}
\end{equation} 
Using equations (\ref{R}) and (\ref{Box}), the $(00)$ and $(ii)$ component of field equation (\ref{GMUNU}) for geometry \ref{metric} read
\begin{equation}
\begin{split}
  & e^{\alpha -\beta} \Bigg[ \Big\{ -2 + 2e^{2\beta} + 4r\beta'\Big\}   + a\Big\{ -e^{2\beta}r^{2}R^{2} -8rR' + 4r^{2}R'\beta'  + 4R( -1 + e^{2\beta} + 2r\alpha') \\& -4r^{2}R'' \Big\} + b \Big\{2U -2e^{2\beta}U   -4rP' + 4rU' -r^{2}P'U' -4rU\beta' + 2r^{2}P'\beta' \\ &-2r^{2}U'\beta' + 2P(-1 + e^{2\beta}  + 2r\beta') -2r^{2}P'' + 2r^{2}U'' \Big\} + c \Big\{ 2S -2e^{2\beta}S - e^{2\beta}r^{2}U^{2} -4rP' + 4rS'  \\& -r^{2}P'U' + r^{2}S'U' -4rS'\beta' + 2r^{2}P'\beta' -2r^{2}S'\beta' + 2P(-1 + e^{2\beta} + 2r\beta') -2r^{2}P'' + 2r^{2}S'' \Big\} \Bigg] =  -\rho  \label{g00}
\end{split}
\end{equation}

and 

\begin{equation}
\begin{split}
 & e^{\alpha -\beta} \Bigg[ \Big\{ -2 + 2e^{2\beta} + 4r\alpha'\Big\}   + a\Big\{ -e^{2\beta}r^{2}R^{2} -8rR' - 4r^{2}R'\alpha'  + 4R( -1 + e^{2\beta} - 2r\alpha') \\&  \Big\} + b \Big\{2U -2e^{2\beta}U   -4rP' + 4rU' + r^{2}P'U' + 4rU\alpha' - 2r^{2}P'\alpha' \\ &+2r^{2}U'\alpha' + 2P(-1 + e^{2\beta}  - 2r\alpha') \Big\} + c \Big\{ 2S -2e^{2\beta}S - e^{2\beta}r^{2}U^{2} -4rP' + 4rS'  \\& +r^{2}P'U' - r^{2}S'U' + 4rS'\alpha' - 2r^{2}P'\alpha' +2r^{2}S'\alpha' + 2P(-1 + e^{2\beta} - 2r\alpha')  \Big\} \Bigg] =  -p, \label{g11}
\end{split}
\end{equation} 
\hspace{2em} and the constraint equation (\ref{traceqn}) becomes
\begin{equation}
R(r)\{b(P-U) +c(P-S) + 1\} = 6a e^{-2\beta}\{ R(r)'' + (\alpha' -\beta' + \frac{2}{r}) + R'(r)\} + (b-c)(U'-P') -c(U'S') + 2cU^{2}, \label{straceqn}
\end{equation}
and equation of motion for auxilary field equations(\ref{ueom}) -(\ref{peom1}) set as
\begin{eqnarray}
e^{-2\beta}\{U''(r) + (\alpha' -\beta' + \frac{2}{r}) + U'(r)\}& = &-R , \label{sueom} \\
e^{-2\beta}\{S''(r) + (\alpha' -\beta' + \frac{2}{r}) + S'(r)\}& = &-U , \label{sseom} \\
(b + c)e^{-2\beta}\{P''(r) + (\alpha' -\beta' + \frac{2}{r}) + P'(r)\}& = & bR + cU . \label{speom}
\end{eqnarray}

\section{Numerical Solutions}  \label{sec3}

Here we evaluate equations (\ref{g00}) -(\ref{speom}) in the region $a^{-\frac{1}{2}} < r < b^{\frac{-1}{2}} < c^{\frac{-1}{2}}$ numerically. This limit on $r$ is chosen to find initial conditions for perturbative estimates of our solution. To solve numerically we need initial conditions for $ \alpha , \beta, S, U, P.$ Note that for the given limit on $r$ we can take $br, cr^{2} \longrightarrow 0,$ then from equation (\ref{traceqn}) 
\begin{equation}
a \Box R = R ,  \label{sstrR}
\end{equation} 
It is quite clear that trivial solution of equation (\ref{sstrR}) is $ R = 0,$ then equations (\ref{g00}) and (\ref{g11}) in the region outside the source, where $ T_{\mu\nu} = 0$, takes the form
\begin{eqnarray}
\alpha' + \beta' & = &0  \label{einstein00}  \\
1 + e^{-2\beta}[r(\beta' -\alpha') -1] & =& 0. \label{einstein11}
\end{eqnarray}

The other equations (\ref{einstein00}) and (\ref{einstein11}) is same as standard GR form for spherically symmetric metric.

Similarly equation of motion for fields $U, S$ become

\begin{eqnarray}
e^{-2\beta}\{U''(r) + (\alpha' -\beta' + \frac{2}{r}) + U'(r)\}& = & 0 , \label{eiueom} \\
e^{-2\beta}\{S''(r) + (\alpha' -\beta' + \frac{2}{r}) + S'(r)\}& = &-U . \label{eiseom} 
\end{eqnarray}

Solutions of (\ref{einstein00}) and (\ref{einstein11}) are given by
\begin{eqnarray}
\alpha(r) &=& \frac{1}{2} \ln(1 -\frac{r_{s}}{r}), \label{salpha} \\
\beta(r) &=& -\frac{1}{2} \ln(1 -\frac{r_{s}}{r}). \label{sbeta}
\end{eqnarray}

Inserting equations (\ref{salpha}) and (\ref{sbeta}) into equation (\ref{eiueom}) we get

\begin{equation}
U(r) = u_{0} -u_{1} \ln (1 - \frac{r_{s}}{r}) ,
\end{equation}

where $u_{0}$ and $ u_{1}$ are constants to parametrize the solution of homogeneous equation (\ref{ueom}).

Following the green function technique as described in ref\cite{Kehagias:2014sda,Maggiore:2014sia} to find the solutions to the field $U$ and by fixing homogeneous part of the solution, we obtain
the solution at linear order

\begin{equation}
U(r) = -\ln (1 -\frac{r_{s}}{r}) .  \label{uin}
\end{equation}

and now plugging expresion of $\alpha ,\beta ,U$ into equation (\ref{eiseom}) \cite{Maggiore:2014sia}, we get expressions for S(r) as 

\begin{equation}
S(r)=\frac{1}{6 r^{5/2}(r-r_s)^{1/2}}
[3 r_s^3-2 r_s^2 r -r_s r^2+2 (r^3-r_s^3)\log\frac{r-r_s}{r_s}
-2 r^3 \log\frac{r}{r_s}] ,  \label{sin}
\end{equation}

Now equipped with equations (\ref{g00}) - (\ref{speom}), we solve for $ e^{2\alpha}$, $ e^{2\beta}$ and scalar auxilary fields U, S, P numerically. Intial values for $ \alpha, \beta, U, S, P$ are obtained by fixing the lower limit of r to $ 200 $. Here we choose $ r = 1 $. We also choose $ a = 10^{-8}$ , $ b = 10^{-7}$ and $ c = 10^{-6.5}$ such that $br, cr < 1. $

\begin{figure}[!ht]
  \centering
 \includegraphics[scale=0.4]{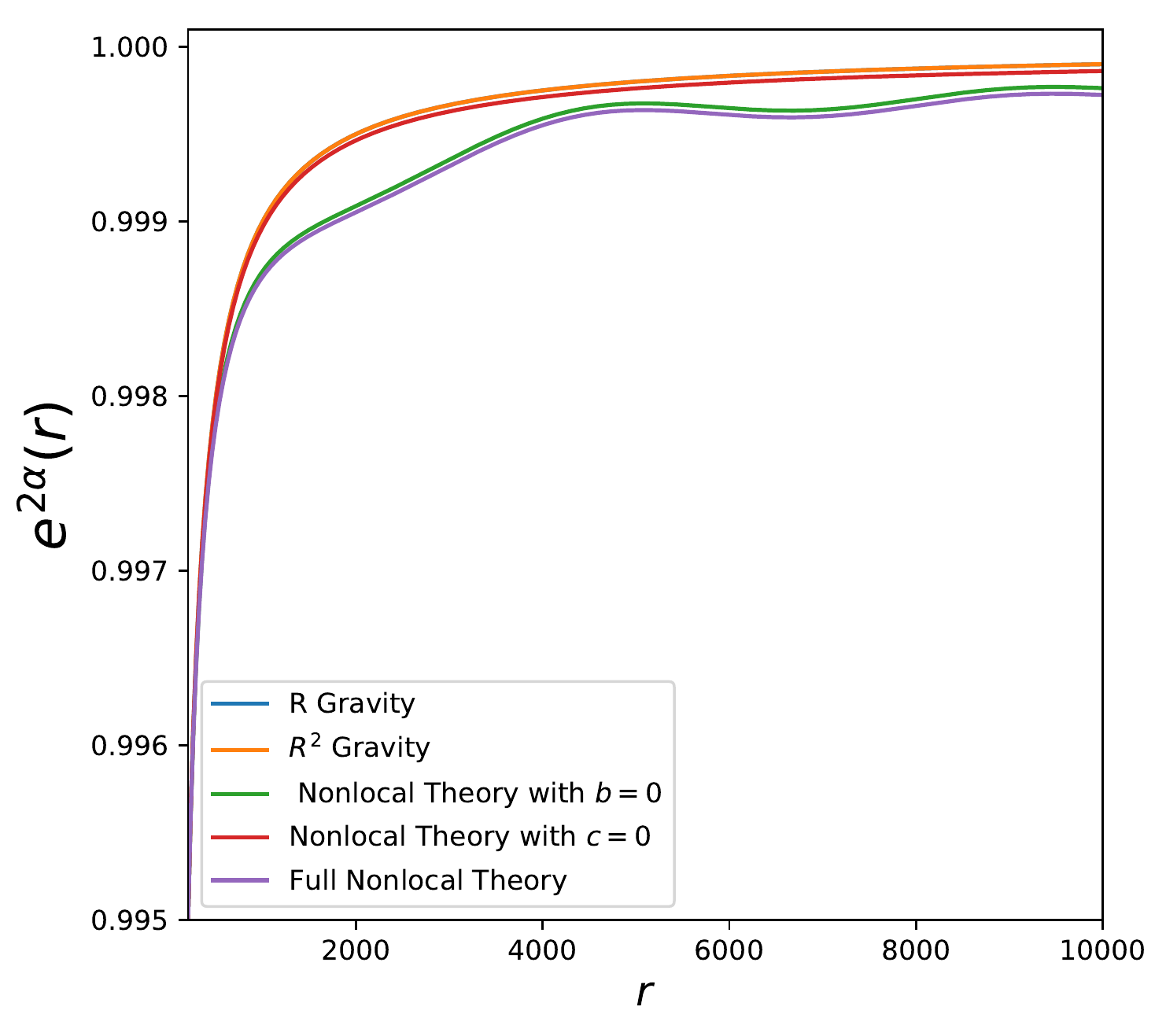} 
 \includegraphics[scale=0.4]{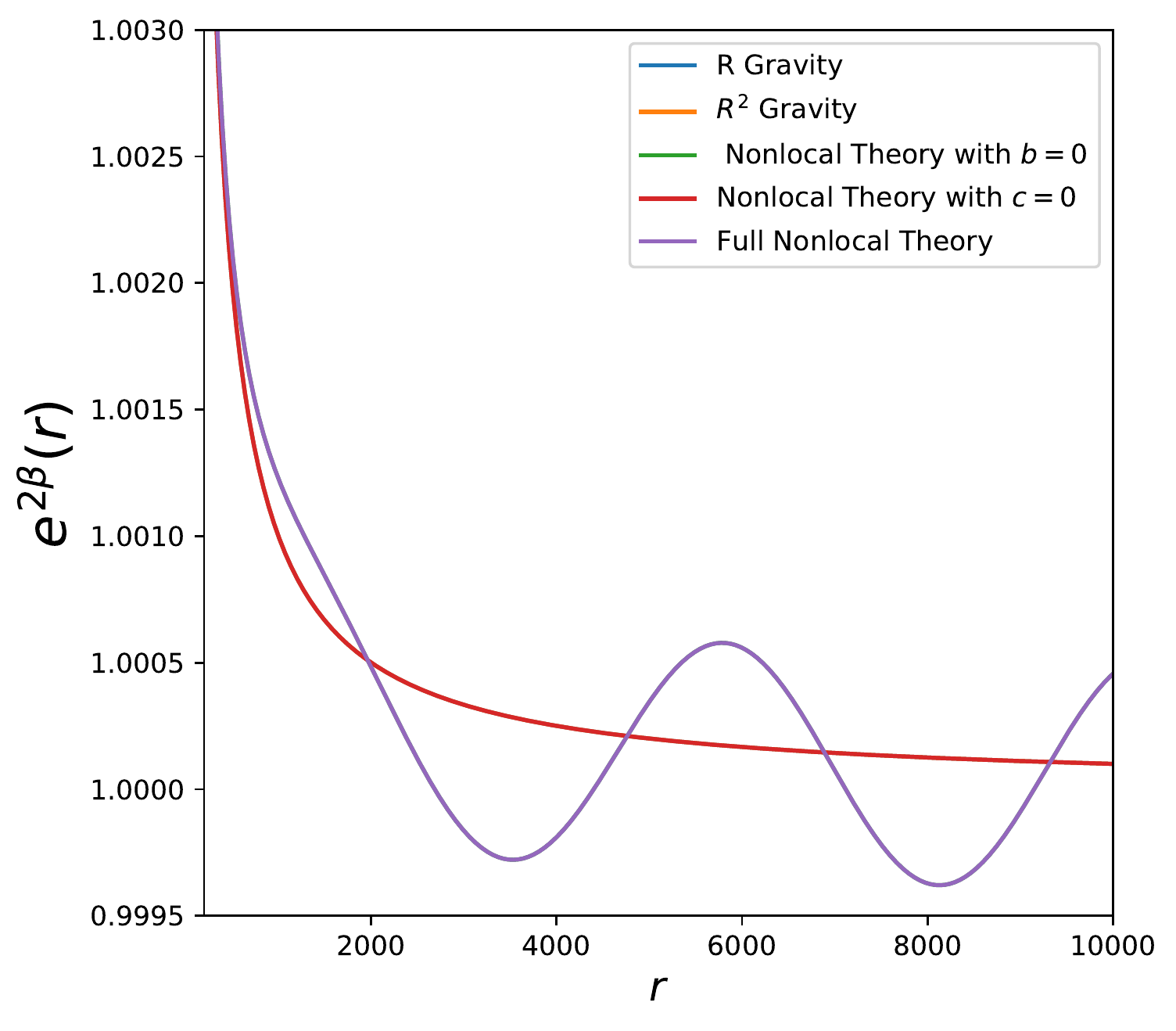} 
\caption{Plots for $e^{2\alpha(r)}$ and $ e^{2\beta(r)}$ for different cases of a,b,c. (colors online).  } 
\label{pab} 
\end{figure}

 Left hand plot of Figure (\ref{pab}) shows the dependence of $A(r) = e^{2\alpha(r)}$ with r, while right hand plot of Figure (\ref{pab}) demonstrates the behaviour of $B(r) = e^{2\beta(r)} $ with $r.$ The plots of $ A(r) $ and $ B(r)$ for  $ R $ and $ R + R^{2}$ overlap completely each other. We also plot $ A(r),B(r)$ for cases $ b = 0$ and $ c = 0 $ which are represented in Figure (\ref{pab}) with green and red lines (online).

\begin{figure}[!ht]
  \centering
 \includegraphics[scale=0.4]{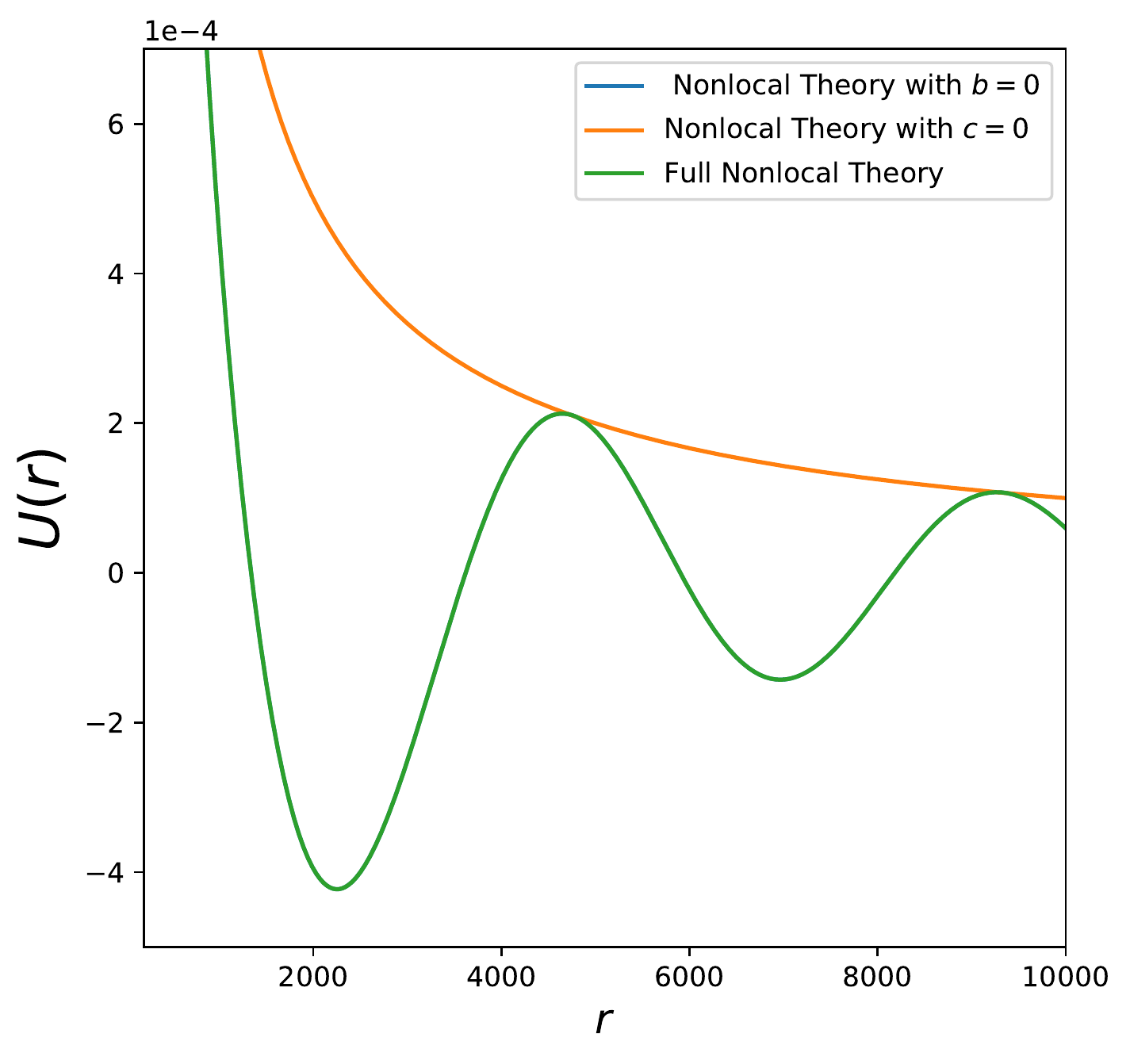} 
 \includegraphics[scale=0.4]{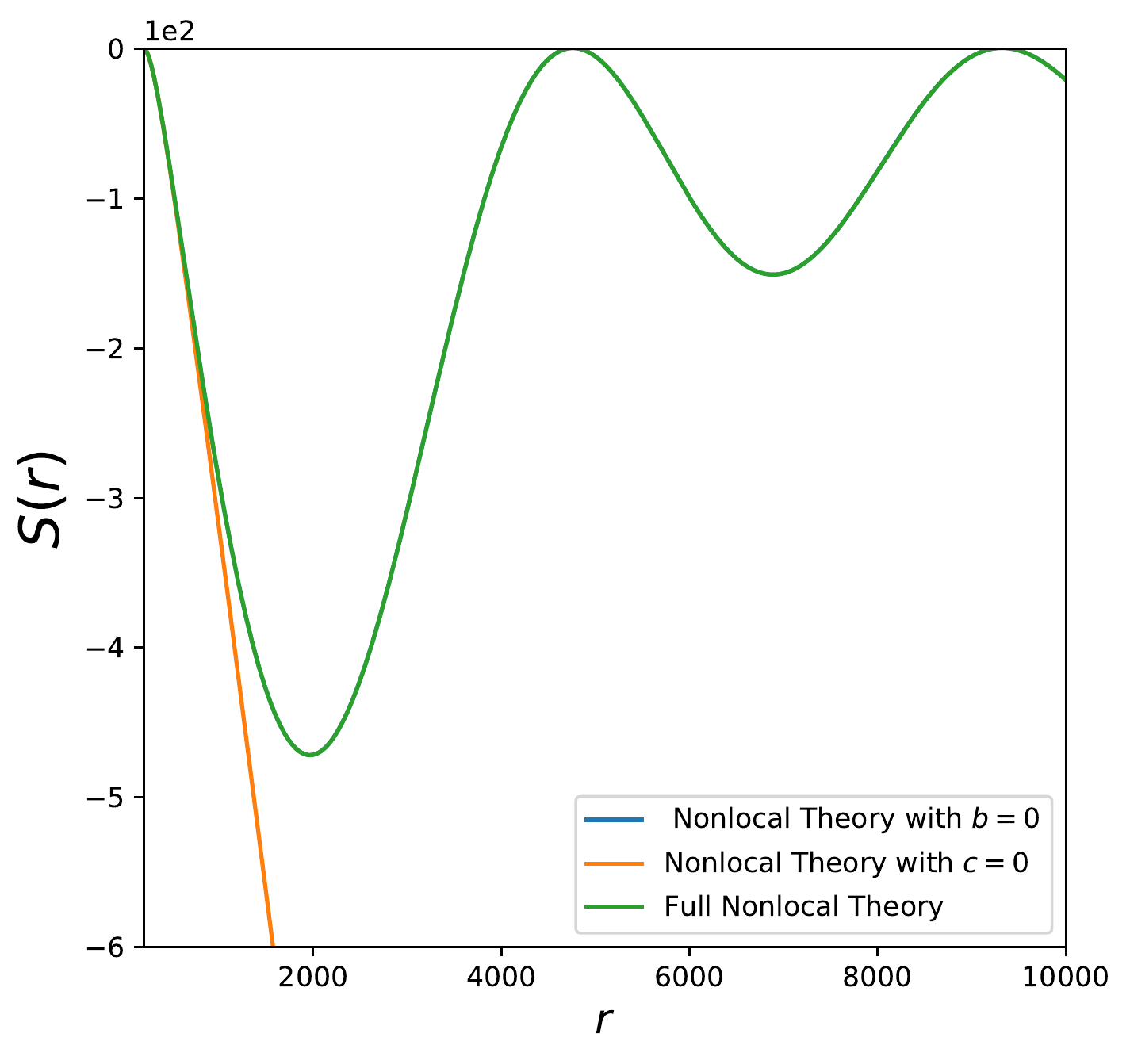} 
\caption{Plots for auxilary scalar fields U(r) and S(r) for different cases of a,b,c (colors online).  } 
\label{pus} 
\end{figure}

The numerical solution for fields U and S shown in Figure (\ref{pus}) for different choices of parameters. 

\begin{figure}[!ht]
  \centering
 \includegraphics[scale=0.4]{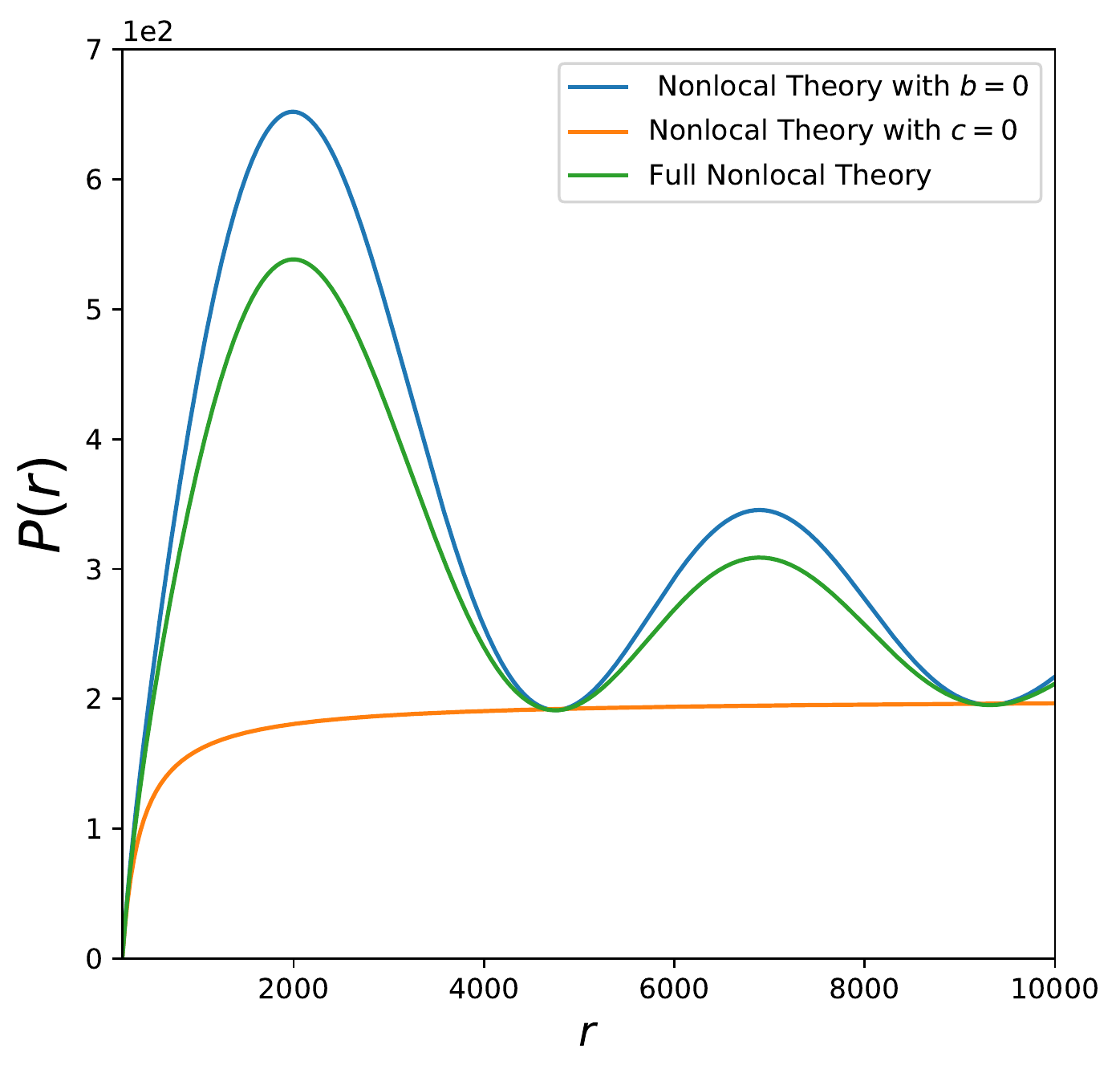} 
 \includegraphics[scale=0.4]{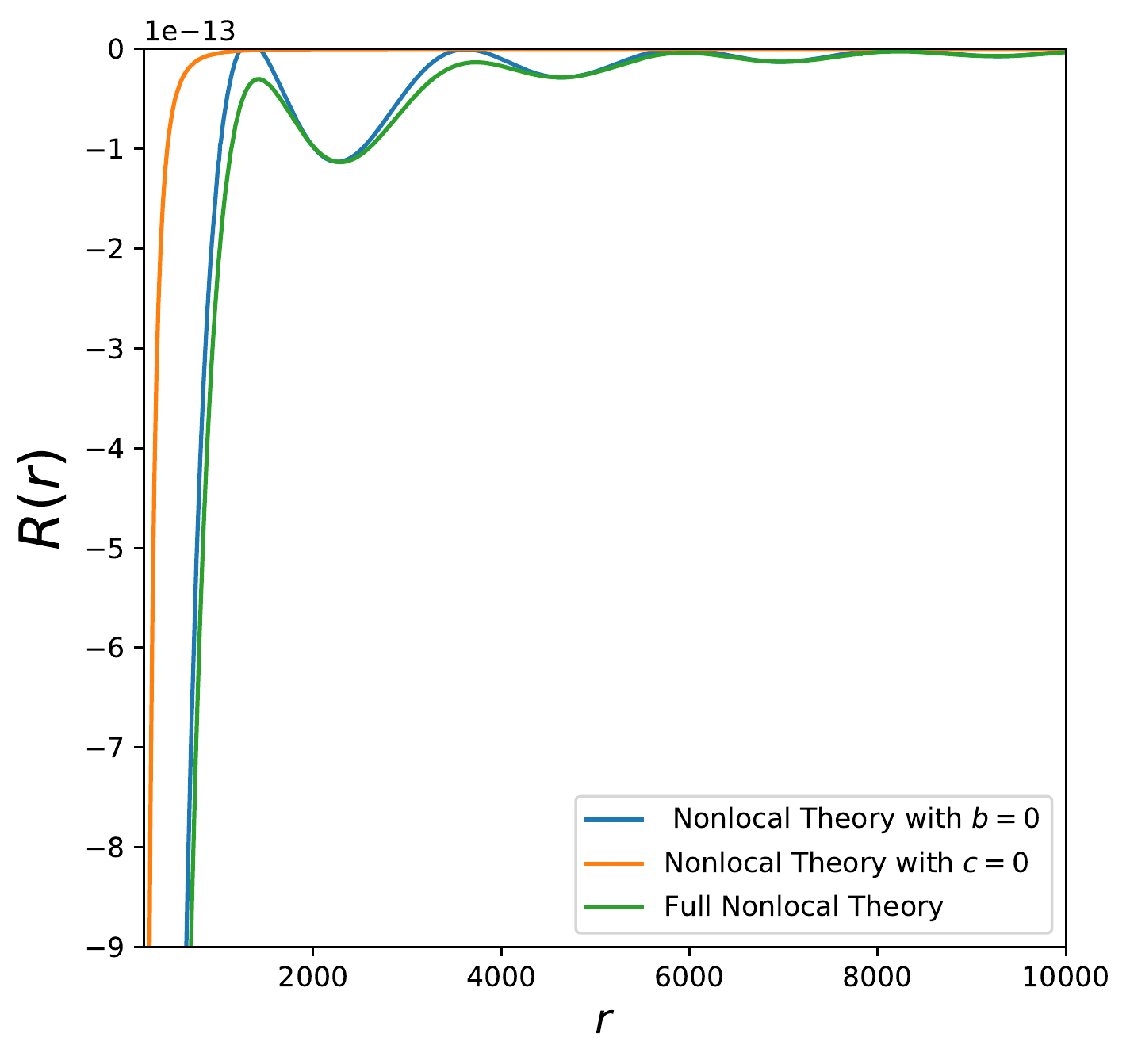} 
\caption{Plots for auxilary scalar field P(r) and Ricci scalar R(r) for different cases of a,b,c (colors online).  } \label{pqp} 
\end{figure}

  Left hand plot of Figure (\ref{pqp}) shows the the numerical solution for field P(r) for various cases. Variation of Ricci scalar R(r) is demonstrated in right hand side plot of Figure (\ref{pqp}). From numerical results it is clear that for 
Einstein gravity and $R^{2}$ gravity ricci scalar becomes 0, while in the nonlocal gravity case it varies very slowly with respect to $r.$ The fields $U, S, P, R$ are oscillatory in nature due to the presence of the nonlocal terms in action. It seems that this is a generic feature of nonlocal models.

\section{Weak field analysis}  \label{sec4} 

\subsection{Field equations: Revisit} In this section we work with original modified field equations for action (\ref{action}) but not with the equivalent scalar tensor action. We follow the method used in ref \cite{Conroy:2014eja}. Variation of the inverse of d'alembertian is given by
\begin{equation}
\delta(\Box^{-1}) S = -\Box^{-1}\delta(\Box)\Box^{-1}.S
\end{equation}

Generalising this to $ \Box^{-n}$ we get
\begin{eqnarray}
\delta(\Box^{-n})S &=& \sum_{m = 0}^{n-1} \Box^{-m}\delta(\Box^{-1})\Box^{-n+m+1}S, \\
&=&  -\sum_{m = 0}^{n-1} \Box^{-m}(\Box^{-1}\delta(\Box)\Box^{-1})\Box^{-n+m+1}S. 
\label{boxvary}
\end{eqnarray} 

Varying the (\ref{action}) with respect to metric $g_{\mu\nu}$ and using equations \ref{boxvary}, we obtain
\begin{equation}
\begin{split}
G_{\mu\nu} + & 2a \Big\{ R\Big( G_{\mu\nu} + \frac{1}{4} R g_{\mu\nu} \Big) + \Big(g_{\mu\nu} \Box R -\nabla_{\mu}\nabla_{\nu}R \Big) \Big\} -2\Big(\nabla_{\mu}\nabla_{\nu} -g_{\mu\nu} \Box  \Big)\Big( b \Box^{-1}-c \Box^{-2}\Big)R  \\ & + \frac{1}{2} g_{\mu\nu}\Big( bR \frac{1}{\Box} R -c R \frac{1}{\Box^{2}}R\Big) +b \nabla_{\mu}R^{-1} \nabla_{\nu}R^{-1} -c\Big( \nabla_{(\mu}R^{-1}\nabla_{\nu)}R^{-2}\Big)  \\ &-\frac{1}{2} g_{\mu\nu}\Big( b \nabla_{\sigma}R^{-1}\nabla^{\sigma}R^{-1} -2c \nabla_{\sigma}R^{-2}\nabla^{\sigma}R^{-1} + b R^{-1} + -2c R^{-2}\Big) = 2\kappa^{2} T_{\mu\nu}.   \label{regmumu}
\end{split}
\end{equation}
 Conserved stress energy tensor $T_{\mu\nu}$ satisfies 
\begin{equation}
\nabla^\mu T_{\mu\nu} = 0.
\end{equation}

\subsection{Weak field limit} 
In order to find the Newtonian limit of our model, we consider the weak field expansion around flat metric of the field equation(\ref{regmumu}). The perturbed metric $g_{\mu\nu} $ around Minkowski metric $ \eta_{\mu\nu},$ is
\begin{equation}
g_{\mu\nu} = \eta_{\mu\nu} + h_{\mu\nu}, \label{pergmu}
\end{equation}
here $ h_{\mu\nu}$ is perturbation around metric $\eta_{\mu\nu}.$ We raise or lower indices by using flat metric  $\eta_{\mu\nu}.$ Using equation(\ref{pergmu}),  Riemann Tensor , Ricci Tensor and Ricci scalar upto linear order in $ h_{\mu\nu}$ are
\begin{eqnarray}
 R_{\rho\mu\sigma\nu} &=& \frac{1}{2} \Big( \partial_{\sigma}\partial_{\mu} h_{\rho\nu} + \partial_{\nu}\partial_{\rho} h_{\mu\sigma} -\partial_{\nu}\partial_{\mu} h_{\rho\sigma} -\partial_{\sigma}\partial_{\rho} h_{\sigma\mu}\Big) , \label{weakrt} \\
R_{\mu\nu}& =& \frac{1}{2}\Big( \partial^{\sigma}\partial_{\mu}h_{\sigma \nu} + \partial_{\nu}\partial_{\sigma} h_{\mu}^{\sigma} -\partial_{\nu}\partial_{\mu}h -\Box h_{\mu\nu} \Big) ,  \label{weakri} \\ 
R &= &\partial_{\mu}\partial_{\nu}h^{\mu\nu} - \Box h . \label{weakrs}
\end{eqnarray}

Using equations (\ref{pergmu})- (\ref{weakrs}) we write field equation (\ref{regmumu}) as 
\begin{equation}
\begin{split}
2\kappa^2 T_{\mu\nu} & =-\Box h_{\mu\nu} + \partial_{\sigma} \partial_{(\mu}h^{\sigma}_{\nu)} + 2a \Big( \partial_{\mu}\partial_{\nu} \partial^{\alpha}\partial^{\beta}\eta_{\alpha \beta} \partial_{\mu}\partial_{\nu}(\Box h) -\eta_{\mu\nu}\Box(\partial_{\mu}\partial_{\nu} -\Box h)\Big) \\& -\Big(1 -2b  + 2c \Box{-1}\Big)(\partial_{\mu}\partial_{\nu}h + \eta_{\mu\nu}\partial_{\alpha}\partial_{\beta}h^{\alpha\beta}) -\Big( -1 +2b -2c\Box^{-1}\Big) \eta_{\mu\nu} \Box h \\& -\Big( 2b -2c\Box^{-1}\Big)\Box^{-1}\nabla_{\mu}\nabla_{\nu}\partial_{\alpha}\partial_{\beta}h^{\alpha\beta}.    \label{wegmunu}
\end{split}
\end{equation}
Re-arranging equation (\ref{wegmunu}) we arrive at

\begin{equation}
\begin{split}
2\kappa^2 T_{\mu\nu} &= -\Bigg[\Box h_{\mu\nu} -\partial_{\sigma} \partial_{(\mu}h^{\sigma}_{\nu)} + \Big(1 -2b  + 2c \Box{-1} -2a\Box \Big)(\partial_{\mu}\partial_{\nu}h + \eta_{\mu\nu}\partial_{\alpha}\partial_{\beta}h^{\alpha\beta}) \\&-\Big( -1 +2b -2c\Box^{-1} + 2a\Box \Big) \eta_{\mu\nu} \Box h  +\Big( 2b -2c\Box^{-1} + 2a\Box \Big)\Box^{-1}\nabla_{\mu}\nabla_{\nu}\partial_{\alpha}\partial_{\beta}h^{\alpha\beta}\Bigg].
\end{split}    
\label{rearrgmunu}
\end{equation}
Now we are ready to use this equation to derive Newtonian potentials.

\subsection{Newtonian Potentials}  \label{sec5}

 By taking the weak field approximation (i.e., $h^{2} \approx 0$) and static limit (i.e., $\Box \approx \nabla^{2}$), the $00$-component and trace of equation(\ref{rearrgmunu}) are

\begin{eqnarray}
-\rho &=& -2(1 - 3b + 3c\Box^{-1} + 3a\Box)(\Box h -\partial_{\alpha}\partial_{\beta}h^{\alpha\beta}), \label{wetr} \\
\rho & = & \Box h_{00} + (1 - 2b +2c\Box^{-1} -2a\Box )(\Box h -\partial_{\alpha}\partial_{\beta}h^{\alpha\beta}). \label{we00}
\end{eqnarray} 

Here we assume $ p \approx 0$ so that trace of $T_{\mu\nu} = -\rho  $ and $ T_{00} = \rho$. Considering a spherically symmetric metric with the line element 
\begin{equation}
ds^2 = -(1 + 2\phi)dt^2 + (1 - 2\psi)dr^2 ,
\end{equation}
and noting that 
\begin{equation}
 h_{00} = -2\phi, \hspace{4 em} h_{ij} = -2\psi \eta_{ij},
\end{equation}
 equations ((\ref{wetr}) and (\ref{we00}) reduces to 
\begin{eqnarray}
-2\kappa^2 \rho &=& -4(1 - 3b +3c\Box^{-1} +3a\Box)(\nabla^{2}\phi -2\nabla^2 \psi), \label{wetrs} \\
2\kappa^2\rho & =  &  -2( -2b + 2c\Box^{-1} - 2a\Box)\nabla^2 \phi - 4(1-2b + 2c\Box^{-1} - 2a\Box) \nabla^2 \psi. \label{we00s}
\end{eqnarray}
Solving equations (\ref{wetrs}) and (\ref{we00s}), expressions for $\nabla^2 \phi$ abd $\nabla^2 \psi$ are
\begin{eqnarray}
\nabla^2 \phi & = & -2\kappa^2 \rho \frac{(-1 + 4b - 4c\Box^{-1} -8a\Box)}{4(-1 +3b -3c\Box^{-1} -3a\Box)} \label{phi} \\
and
\nabla^2 \psi & = & -2\kappa^2 \rho \frac{(-1 + 2b - 2c\Box^{-1} -4a\Box)}{4(-1 +3b -3c\Box^{-1} -3a\Box)}. \label{psi}
\end{eqnarray}

Taking fourier transform of equation (\ref{phi}) and (\ref{psi}) and using $M_{p}^{2} = \frac{1}{2\kappa^{2}}$, $M_p$ is the Plank mass, we express newtonian potential $\phi (r)$ and $\psi(r)$ as
\begin{eqnarray}
\phi(r)& = & -\frac{m}{(4\pi)^3 M_{p}^{2}} \int_{-\infty}^{\infty} d^{3}k \frac{e^{i\vec{k}\vec{r}}(4c -k^{2} + 4bk^{2} + 8ak^{4})}{4(3c -k^2 + 3bk^2 + 3ak^4)} \\ 
&=& -\frac{m}{\pi^2 M_{p}^{2}r} \int_{0}^{\infty} \frac{dk}{k}  \frac{\sin(kr)(4c -k^{2} + 4bk^{2} + 8ak^{4})}{4(3c -k^2 + 3bk^2 + 3ak^4)}, \label{phipoten}
\end{eqnarray}

\begin{eqnarray}
\psi(r)& = & -\frac{m}{(4\pi)^3 M_{p}^{2}} \int_{-\infty}^{\infty} d^{3}k \frac{e^{i\vec{k}\vec{r}}(2c -k^{2} + 2bk^{2} + 4ak^{4})}{4(3c -k^2 + 3bk^2 + 3ak^4)} \\ 
&=& -\frac{m}{\pi^2 M_{p}^{2}r} \int_{0}^{\infty} \frac{dk}{k}  \frac{\sin(kr)(2c -k^{2} + 2bk^{2} + 4ak^{4})}{4(3c -k^2 + 3bk^2 + 3ak^4)}. \label{psipoten}
\end{eqnarray}

By performing contour integration of these intergrals (\ref{phipoten}) and (\ref{psipoten}), we obtain

\begin{eqnarray}
\phi(r) &=& \frac{m}{24 M_{p}^{2}r}\bigg\{\frac{A e^{-M_1 r}}{C} + \frac{B e^{-M_{2}r}}{D} + 4 \bigg\}  \label{ph}\\ 
\psi(r) &=& \frac{m}{24 M_{p}^{2}r}\bigg\{\frac{E e^{-M_1 r}}{C} + \frac{F e^{-M_{2}r}}{D} + 2 \bigg\} \label{ps}
\end{eqnarray}
where $A, B, C, D, E, F,M1,M2 $  are given by
\begin{eqnarray}
A &=&  36b^2 - 72ac +5 + 5\sqrt{1 - 6b + 9b^2 -36ac} - 3b( 9 + 4\sqrt{1-6b + 9b^2 -36ac})  \label{a}\\
B &=&  5 + 36b^2 -72ac -5\sqrt{1 -6b + 9b^2 -36ac} + 3b( -9 + 4\sqrt{1 - 6b + 9b^2 -36ac}) \label{b} \\
C &=&  1 + 9b^2 -36ac + \sqrt{1 -6b + 9b^2 -36ac} -3b(2 + \sqrt{1 - 6b + 9b^2 -36ac})\label{c} \\
D &=&  1 + 9b^2 -36ac -\sqrt{1 - 6b + 9b^2 -36ac} + 3b( -2 + \sqrt{1 -6b + 9b^2 -36ac}) \label{d}\\ 
E &=&  1 + 18b^2 -36ac + \sqrt{1 -6b + 9b^2 -36ac} + b(9 + 6\sqrt{1-6b + 9b^2 -36ac}) \label{e}\\
F &=&  1 + 18b^2 -36ac -\sqrt{1 -6b + 9b^2 -36ac} + b(-9 + 6\sqrt{1 -6b + 9b^2 -36ac}) \label{f}\\
M_{1} &=& \sqrt{-\frac{1-3b + \sqrt{1-6b + 9b^2 -36ac}}{6a}} \label{m1}\\ 
M_{2} &=& \sqrt{\frac{-1+3b + \sqrt{1-6b + 9b^2 -36ac}}{6a}} \label{m2}
\end{eqnarray}

It is clear from equations (\ref{ph}) and (\ref{ps}) that 
$\phi(r) $ and $ \psi(r)$  are free from singularity 
provided $ C$ and $ D$ are nonzero. As we see from equations (\ref{a}) -(\ref{m2}) only real solutions are possible for  
$\phi(r) $ and $ \psi(r)$ if $ 1 - 6b + 9b^2 -36ac > 0 $. For exponentially decaying real solutions we have a further constraint coming from equations (\ref{m1})- (\ref{m2}), i.e., $ 1 - 6b + 9b^2 -36ac >1 .$ 
Attractive and repulsive nature of the potential depends on the value of $A, B, C$ and $F$ in equations (\ref{ph}) and (\ref{ps}). $M_{1}$ and $M_{2}$ suggest the presence of extra massive degrees of freedom in our theory. Next we shall find the exact no. of degrees of freedom for our action.

\section{Counting Degrees of Freedom (dof)} \label{sec6} 

In order to count dof for our nonlocal model (\ref{complteaction}), we derive 
the propagator  for $ h_{\mu\nu} $ field. We follow the algorithm provided in 
Ref. \cite{VanNieuwenhuizen:1973fi,Biswas:2013kla}. 

Linearized field equation (\ref{rearrgmunu}) can be expressed as
\begin{equation}
\Pi_{\mu\nu}^{-1\lambda\sigma} h_{\lambda \sigma} = 2\kappa^{2}T_{\mu\nu},   \label{prop}
\end{equation}
where $ \Pi_{\mu\nu}^{-1\lambda\sigma} $ is the inverse of propagator and can be expressed in terms of projection operators in following way.
\begin{equation}
\Pi_{\mu\nu}^{-1\lambda\sigma} = \sum^{6}_{i = 1}c_{i}\mathcal{P}_{i}, \label{defpro}
\end{equation}
where $ \mathcal{P}_{i}{\hskip 0.5em} $ are spin projector operators. Introducing 6 spin projector operators which forms the complete set are \cite{VanNieuwenhuizen:1973fi,Biswas:2013kla}
\begin{eqnarray}
\mathcal{P}^{2}_{\mu\rho\nu\sigma}& = &\frac{1}{2}(\theta_{\mu\rho}\theta_{\nu\sigma} + \theta_{\mu\sigma}\theta_{\nu\rho}) - \frac{1}{3}\theta_{\mu\nu}\theta_{\rho\sigma} , \label{spin2} \\
\mathcal{P}^{1}_{\mu\rho\nu\sigma}& = &\frac{1}{2}(\theta_{\mu\rho}\omega_{\nu\sigma} + \theta_{\mu\sigma}\omega_{\nu\rho}  + \theta_{\nu\rho}\omega_{\mu\sigma} + \theta_{\nu\sigma}\omega_{\mu\rho}) \label{spin1} \\
\mathcal{P}^{0s}_{\mu\nu\sigma\rho} & = & \frac{1}{3} \theta_{\mu\nu}\theta_{\rho\sigma} \hspace{2 em},\hspace{2 em} \mathcal{P}^{0w}_{\mu\nu\sigma\rho} = \omega_{\mu\nu}\omega_{\rho\sigma} , \label{spin0s} \\
 \mathcal{P}^{0sw}_{\mu\nu\sigma\rho} & = & \frac{1}{\sqrt{3}} \theta_{\mu\nu}\omega_{\rho\sigma} \hspace{2 em},\hspace{2 em} \mathcal{P}^{0ws}_{\mu\nu\sigma\rho} = \omega_{\mu\nu}\theta_{\rho\sigma} , \label{spin0sw}
\end{eqnarray}
where $ \theta_{\mu\nu}$ and $ \omega_{\mu\nu}$ are called transverse and longitudinal operators defined in k space have the following form
\begin{equation}
\theta_{\mu\nu} = \eta_{\mu\nu} - \omega_{\mu\nu} \hspace{2em} , \hspace{2em} \omega_{\mu\nu} = \frac{k_{\mu}k_{\nu}}{k^{2}} .
\end{equation} \\
$\mathcal{P}^{2}_{\mu\rho\nu\sigma}$ and $\mathcal{P}^{1}_{\mu\rho\nu\sigma}$ operators represent transverse and traceless spin-2 and spin-1 degrees, whereas $\mathcal{P}^{0s}_{\mu\nu\sigma\rho}$ and $ \mathcal{P}^{0w}_{\mu\nu\sigma\rho} $  represent the spin-0 scalar multipltes. In addition to these four spin degrees we also have two more scalar multiplets $ \mathcal{P}^{0sw}_{\mu\nu\sigma\rho}$ and $ \mathcal{P}^{0ws}_{\mu\nu\sigma\rho}$.

Using equation (\ref{defpro}), we can rewrite equation(\ref{prop}) as
\begin{equation}
\sum^{6}_{i = 1}c_{i}\mathcal{P_{\mu\rho\sigma\nu}}_{i}h = 2\kappa^{2}(\mathcal{P}^{2}_{\mu\rho\nu\sigma} + \mathcal{P}^{1}_{\mu\rho\nu\sigma} + \mathcal{P}^{0s}_{\mu\rho\nu\sigma} + \mathcal{P}^{0w}_{\mu\rho\nu\sigma})T,
\end{equation} 
where $i$ takes the all projector operators described above.
\subsection{Inverting Field equations}
Using projector operators, Each term in equation (\ref{rearrgmunu}) becomes
\begin{eqnarray}
\Box h_{\mu\nu} &=& -k^2 \Big[\mathcal{P}^{2}_{\mu\rho\nu\sigma} + \mathcal{P}^{1}_{\mu\rho\nu\sigma} + \mathcal{P}^{0s}_{\mu\rho\nu\sigma} + \mathcal{P}^{0w}_{\mu\rho\nu\sigma}\Big]h \label{1tp} \\ 
\partial_{\sigma} \partial_{(\mu}h^{\sigma}_{\nu)} &=& -k^2\Big[ \mathcal{P}^{1}_{\mu\rho\nu\sigma} +2\mathcal{P}^{0w}_{\mu\rho\nu\sigma} \Big]h \label{2tp} \\
d(\Box)(\partial_{\mu}\partial_{\nu}h + \eta_{\mu\nu}\partial_{\alpha}\partial_{\beta}h^{\alpha\beta}) & = & -d(-k^{2})k^{2}\Big[ 2\mathcal{P}^{0w}_{\mu\rho\nu\sigma} + \sqrt{3}\Big( \mathcal{P}^{0sw}_{\mu\rho\nu\sigma} + \mathcal{P}^{0ws}_{\mu\rho\nu\sigma}\Big)\Big]h \label{3tp} \\
\eta_{\mu\nu} e(\Box)\Box h & = &-k^{2}d(-k^{2})\Big[3\mathcal{P}^{0s}_{\mu\rho\nu\sigma}  + \mathcal{P}^{0w}_{\mu\rho\nu\sigma} + \sqrt{3}\Big( \mathcal{P}^{0sw}_{\mu\rho\nu\sigma} + \mathcal{P}^{0ws}_{\mu\rho\nu\sigma}  \Big) \Big]h,  \label{4tp} \\
f(\Box)\Box^{-1}\nabla_{\mu}\nabla_{\nu}\partial_{\alpha}\partial_{\beta}h^{\alpha\beta} &=&  -f(-k^{2})k^2 \mathcal{P}^{0w}_{\mu\rho\nu\sigma}  \label{5tp}
\end{eqnarray} 
where  we define $d(\Box) = \Big(1 -2b  + 2c \Box{-1} -2a\Box \Big) $, $e(\Box) = -\Big(1 -2b  + 2c \Box{-1} -2a\Box \Big) $ and $f(\Box) =  \Box^{-1}\nabla_{\mu}\nabla_{\nu}\partial_{\alpha}\partial_{\beta}h^{\alpha\beta}.$
{\vskip 2em}

Using equations (\ref{1tp})- (\ref{5tp}), equation (\ref{prop}) becomes
\begin{equation}
\begin{split}
2\kappa^{2}(\mathcal{P}^{2}_{\mu\rho\nu\sigma} + \mathcal{P}^{1}_{\mu\rho\nu\sigma} + \mathcal{P}^{0s}_{\mu\rho\nu\sigma} + \mathcal{P}^{0w}_{\mu\rho\nu\sigma})T  = &
\mathcal{P}^{2}_{\mu\rho\nu\sigma} + \mathcal{P}^{0w}_{\mu\rho\nu\sigma}(-k^2 + 
2d(-k^2)k^2 + e(-k^2)k^2)  \\&+ \mathcal{P}^{0s}_{\mu\rho\nu\sigma}(1 + 3e(-k^2))  + \sqrt{3}k^{2}\mathcal{P}^{0sw}_{\mu\rho\nu\sigma}(d(-k^2) \\& + e(-k^2)) + \sqrt{3}k^{2}\mathcal{P}^{0wS}_{\mu\rho\nu\sigma}(d(-k^2) + e(-k^2))h  \label{finalprop2}
\end{split}
\end{equation}
Projecting spin operator $\mathcal{P}^{2}_{\mu\rho\nu\sigma},\mathcal{P}^{1}_{\mu\rho\nu\sigma}$ upon equation (\ref{finalprop2}) we find 
\begin{eqnarray}
\mathcal{P}^{2}_{\mu\rho\nu\sigma}h &=& 2\kappa^{2}\Bigg(\frac{\mathcal{P}^{2}_{\mu\rho\nu\sigma}}{k^{2}}\Bigg) T \label{prop1} \\
\mathcal{P}^{1}_{\mu\rho\nu\sigma} &=& 0. \label{prop2} 
\end{eqnarray}
Action of $\mathcal{P}^{0s}_{\mu\rho\nu\sigma} $ and $\mathcal{P}^{0w}_{\mu\rho\nu\sigma} $ on equation (\ref{finalprop2}) we get
\begin{eqnarray}
 (1 + 3e) \mathcal{P}^{0s}_{\mu\rho\nu\sigma}hk^2 + \sqrt{3}(d + e)\mathcal{P}^{0sw}_{\mu\rho\nu\sigma} k^{2}h &=& 2\kappa^2 \mathcal{P}^{0s}_{\mu\rho\nu\sigma} T \label{prop3} \\
\sqrt{3}(d + e)\mathcal{P}^{0ws}_{\mu\rho\nu\sigma} k^2 h + (-1 + 2d + e + f)\mathcal{P}^{0w}_{\mu\rho\nu\sigma} k^2 h &=& 2\kappa^{2}T \label{prop4} 
\end{eqnarray}
It is clear from equations (\ref{prop3}) and (\ref{prop4}) that scalar multiplets are coupled. To decouple these multiplets from each other, we project again $ \mathcal{P}^{0w}_{\mu\rho\nu\sigma} $  and $\mathcal{P}^{0s}_{\mu\rho\nu\sigma}$ upon equations (\ref{prop3}) and (\ref{prop4}) respectively, we get
\begin{eqnarray}
 \mathcal{P}^{0s}_{\mu\rho\nu\sigma}h &=& 2\kappa^{2}\frac{\mathcal{P}^{0s}_{\mu\rho\nu\sigma}}{(1 + 3e)k^{2}}T , \label{prop3f} \\
 \mathcal{P}^{0w}_{\mu\rho\nu\sigma} &=&  2\kappa^{2}\frac{\mathcal{P}^{0s}_{\mu\rho\nu\sigma}}{(-1 + 2d + e + f)k^{2}}T.   \label{prop4f}
\end{eqnarray}
Note that the denominator of equation (\ref{prop4f}) becomes 0. Finally total propagator we obtain
\begin{equation}
\Pi_{\mu\rho\nu\sigma} = \frac{\mathcal{P}^{2}_{\mu\rho\nu\sigma}}{k^2} + \frac{\mathcal{P}^{0s}_{\mu\rho\nu\sigma}}{k^{2}(1+3e)}   \label{genprop}
\end{equation}
Further simplifying 
\begin{equation}
\Pi_{\mu\rho\nu\sigma} = \frac{\mathcal{P}^{2}_{\mu\rho\nu\sigma}}{k^2} - \frac{\mathcal{P}^{0s}_{\mu\rho\nu\sigma}}{( 6ak^{4} + k^{2}(2-6b) -6c )} \label{finalprop1}
\end{equation}

Now we are in a position to count the degrees of freedom of our non-local theories by inspecting the propagator in equation (\ref{finalprop1}). We can get the number of dof by looking at the poles of the propagator. Poles of equation (\ref{finalprop1}) are given by  solutions of following equations 
\begin{eqnarray}
 k^2 &=& 0  \label{1dof} \\
 (6ak^{4} + (2-6b)k^{2} -6c) &=& 0. \label{2dof}
\end{eqnarray}
From equations (\ref{1dof}) and (\ref{2dof}) that we have total three degrees of freedom. One of them is massless of spin 2  multiplet   while other two are the massive degrees of freedom of spin 0 singlet. Mass of these gravitions can be found by solving equation (\ref{2dof}). These masses are same as in expression of $\phi(r)$ and $\psi(r) $ in section \ref{sec5}. Now we will study the role of individual terms of our action.
\section{Special Cases} \label{sec7}
Here we consider some examples by taking different values of a, b, c.
\subsection{Case 1 : RR Model}  \label{case 1potent}
\subsubsection{Newtonian Potential and dof}
Taking $ a = 0$ and $ b= 0$ then our action reduces to RR nolocal model \cite{Maggiore:2014sia}. We find the expression for newtonian potential $ \phi(r)$ as 
\begin{equation}
\phi(r) = -\frac{m}{24 \pi M_{p}^{2}r}\Big[ e^{-M_{1}r} + 8 \Big]. \label{rrphi}
\end{equation}
Similarly exprssion for $\psi(r)$ is given by
\begin{equation}
\psi(r) = \frac{m}{24 \pi M_{p}^{2}r}\Big[ e^{-M_{1}r} - 4\Big].    \label{rrpsi}
\end{equation}
 Propagator for RR model as
\begin{equation}
\Pi = \frac{\mathcal{P}^{2}_{\mu\rho\nu\sigma}}{k^2} - \frac{\mathcal{P}^{0s}_{\mu\rho\nu\sigma}}{(2 k^{2} -6c )} \label{finalpropcase1}
\end{equation}

In this case we have total 2 degrees of freedom, a massless spin 2 multiplet  while the other one is a massive spin 0 singlet.
\subsection{Case 2 : STAROBINSKY MODEL } \label{case 2potent}
\subsubsection{Newtonian Potential and dof}
Starobinsky model \cite{Yu:2017uyd} is given by the action
\begin{equation}
S = \int d^{4}x \sqrt{-g} \Big[  R + a R^2 +  2 \kappa^2\mathcal{L}_{m} \Big].
\end{equation}
Here $ b = c = 0$, in this particular case we obtain $ \phi(r)$ and $ \psi(r)$ as 
\begin{eqnarray}
\phi(r) &=& -\frac{m}{24 \pi M_{p}^{2}r}\Big[ 5 e^{-M_{1}r} + 6 \Big]. \label{stphi}  \\
\psi(r) &=& -\frac{m}{24 \pi M_{p}^{2}r}\Big[ e^{-M_{2}r}  + 6\Big].   \label{stpsi}
\end{eqnarray}
 Propagator for starobinsky model as
\begin{equation}
\Pi = \frac{\mathcal{P}^{2}_{\mu\rho\nu\sigma}}{k^2} - \frac{\mathcal{P}^{0s}_{\mu\rho\nu\sigma}}{2k^{2}(1 + 3ak^{2})} \label{finalpropcase2}
\end{equation}
In this case we have total 3 degrees of freedom, two massless with spin 2 multiplet and spin 0 singlet, third one is spin 0 singlet with mass $\sqrt{\frac{1}{-3a}}$. Mass becomes real if $a < 0 $.

\subsection{Case 3 : Deser Woodard Model with $ f(\frac{R}{\Box}) = \frac{R}{\Box}$}
\subsubsection{Newtonian Potential and dof} For $ a = 0$ and $ c = 0 $, we get the Deser and woodard model  for  $ f(\frac{R}{\Box}) = \frac{R}{\Box}$\cite{Deser:2007jk}. In this case, Newtonian potentials as obtained in equation (\ref{ph}) and (\ref{ps}) become
\begin{eqnarray}
\phi(r) &=& -\frac{m}{4 \pi M_{p}^{2}r}\Bigg[ \frac{4b -1}{3b-1}\Bigg], \label{dephi}  \\
\psi(r) &=& -\frac{m}{4 \pi M_{p}^{2}r}\Bigg[ \frac{2b -1}{3b-1}\Bigg].   \label{depsi}
\end{eqnarray}

 Propagator for this model is
\begin{equation}
\Pi = \frac{\mathcal{P}^{2}_{\mu\rho\nu\sigma}}{k^2} - \frac{\mathcal{P}^{0s}_{\mu\rho\nu\sigma}}{k^{2}(2-6b)} \label{finalpropcase3}
\end{equation}
 In this case we have total 2 degrees of freedom, both of them are massless (one is spin 2 multiplet where as other one is spin 0 singlet)
\section{Summary} \label{sec8}
In this paper, we have presented the spherically symmetric static solutions for a non local action inspired by effective theory of quantum gravity. In addition we have calculated newtonian potentials by taking weak graviational field expansion for our non local action.
We derive the graviton propagator and obtain the number of degrees of freedom
present for our nonlocal action. It turns out that we have one massless spin 2 multiplet and two massive spin 0 singlet for our model. Finally, for different limiting cases of our action related to well known cosmological models we count the degrees of freedom.  In future we will extend our analysis to other nonlocal models.

\section{Acknowledgement} This work was partially funded by DST grant no. SERB/PHY/2017041.


\newpage
\begin{thebibliography}{99}
  \bibitem{Riess:1998cb}
  A.~G.~Riess {\it et al.} [Supernova Search Team],
  Astron.\ J.\  {\bf 116} (1998) 1009
  doi:10.1086/300499
  [astro-ph/9805201].
\bibitem{Rubin:2016iqe}
  D.~Rubin and B.~Hayden,
  Astrophys.\ J.\  {\bf 833} (2016) no.2,  L30
  doi:10.3847/2041-8213/833/2/L30
  [arXiv:1610.08972 [astro-ph.CO]].
  \bibitem{Deser:2007jk}
  S.~Deser and R.~P.~Woodard,
  Phys.\ Rev.\ Lett.\  {\bf 99} (2007) 111301
  doi:10.1103/PhysRevLett.99.111301
  [arXiv:0706.2151 [astro-ph]].
  \bibitem{Woodard:2014iga}
  R.~P.~Woodard,
  Found.\ Phys.\  {\bf 44} (2014) 213
  doi:10.1007/s10701-014-9780-6
  [arXiv:1401.0254 [astro-ph.CO]].
  \bibitem{Koivisto:2008dh}
  T.~S.~Koivisto,
  Phys.\ Rev.\ D {\bf 78} (2008) 123505
  doi:10.1103/PhysRevD.78.123505
  [arXiv:0807.3778 [gr-qc]].
  \bibitem{Dodelson:2013sma}
  S.~Dodelson and S.~Park,
  Phys.\ Rev.\ D {\bf 90} (2014) 043535
  doi:10.1103/PhysRevD.90.043535
  [arXiv:1310.4329 [astro-ph.CO]].
  \bibitem{Dirian:2014xoa}
  Y.~Dirian and E.~Mitsou,
  JCAP {\bf 1410} (2014) no.10,  065
  doi:10.1088/1475-7516/2014/10/065
  [arXiv:1408.5058 [gr-qc]].
  \bibitem{Nersisyan:2016hjh}
  H.~Nersisyan, Y.~Akrami, L.~Amendola, T.~S.~Koivisto and J.~Rubio,
  Phys.\ Rev.\ D {\bf 94} (2016) no.4,  043531
  doi:10.1103/PhysRevD.94.043531
  [arXiv:1606.04349 [gr-qc]].
  \bibitem{Cusin:2015rex}
  G.~Cusin, S.~Foffa, M.~Maggiore and M.~Mancarella,
  Phys.\ Rev.\ D {\bf 93} (2016) no.4,  043006
  doi:10.1103/PhysRevD.93.043006
  [arXiv:1512.06373 [hep-th]].
  \bibitem{Tsamis:2009ja}
  N.~C.~Tsamis and R.~P.~Woodard,
  Phys.\ Rev.\ D {\bf 80} (2009) 083512
  doi:10.1103/PhysRevD.80.083512
  [arXiv:0904.2368 [gr-qc]].
  \bibitem{Tsamis:2010ph}
  N.~C.~Tsamis and R.~P.~Woodard,
  Phys.\ Rev.\ D {\bf 81} (2010) 103509
  doi:10.1103/PhysRevD.81.103509
  [arXiv:1001.4929 [gr-qc]].
\bibitem{Romania:2012av} 
  M.~G.~Romania, N.~C.~Tsamis and R.~P.~Woodard,
  Lect.\ Notes Phys.\  {\bf 863}, 375 (2013)
  doi:10.1007/978-3-642-33036-0-13
  [arXiv:1204.6558 [gr-qc]].
\bibitem{Codello:2016neo}
  A.~Codello and R.~K.~Jain,
  Eur.\ Phys.\ J.\ C {\bf 78} (2018) no.5,  357
  doi:10.1140/epjc/s10052-018-5839-4
  [arXiv:1603.00028 [gr-qc]].
\bibitem{Codello:2016xhm}
  A.~Codello and R.~K.~Jain,
  Int.\ J.\ Mod.\ Phys.\ D {\bf 25} (2016) no.12,  1644023
  doi:10.1142/S0218271816440235
  [arXiv:1605.07630 [gr-qc]]. 
\bibitem{Belgacem:2017cqo}
  E.~Belgacem, Y.~Dirian, S.~Foffa and M.~Maggiore,
  JCAP {\bf 1803} (2018) no.03,  002
  doi:10.1088/1475-7516/2018/03/002
  [arXiv:1712.07066 [hep-th]].
  \bibitem{Park:2017zls}
  S.~Park,
  Phys.\ Rev.\ D {\bf 97} (2018) no.4,  044006
  doi:10.1103/PhysRevD.97.044006
  [arXiv:1711.08759 [gr-qc]].
  \bibitem{Kumar:2018pkb}
  U.~Kumar, S.~Panda and A.~Patel,
  arXiv:1808.04569 [gr-qc].
  \bibitem{Kumar:2018chy}
  U.~Kumar and S.~Panda,
  arXiv:1806.09616 [gr-qc].
  \bibitem{Amendola:2017qge}
  L.~Amendola, N.~Burzilla and H.~Nersisyan,
  Phys.\ Rev.\ D {\bf 96} (2017) no.8,  084031
  doi:10.1103/PhysRevD.96.084031
  [arXiv:1707.04628 [gr-qc]].  
  \bibitem{Codello:2016elq}
  A.~Codello and R.~K.~Jain,
  PoS DSU {\bf 2015} (2016) 008.
  doi:10.22323/1.268.0008
  \bibitem{Codello:2015pga}
  A.~Codello and R.~K.~Jain,
  Class.\ Quant.\ Grav.\  {\bf 34} (2017) no.3,  035015
  doi:10.1088/1361-6382/aa549d
  [arXiv:1507.07829 [astro-ph.CO]].
  \bibitem{Maggiore:2016fbn}
  M.~Maggiore,
  Phys.\ Rev.\ D {\bf 93} (2016) no.6,  063008
  doi:10.1103/PhysRevD.93.063008
  [arXiv:1603.01515 [hep-th]].
  \bibitem{Donoghue:1993eb}
  J.~F.~Donoghue,
  Phys.\ Rev.\ Lett.\  {\bf 72} (1994) 2996
  doi:10.1103/PhysRevLett.72.2996
  [gr-qc/9310024].
  \bibitem{Donoghue:1994dn}
  J.~F.~Donoghue,
  Phys.\ Rev.\ D {\bf 50} (1994) 3874
  doi:10.1103/PhysRevD.50.3874
  [gr-qc/9405057].
  \bibitem{Starobinsky:1980te}
  A.~A.~Starobinsky,
  Phys.\ Lett.\ B {\bf 91} (1980) 99
   [Phys.\ Lett.\  {\bf 91B} (1980) 99].
  doi:10.1016/0370-2693(80)90670-X
  \bibitem{Mukhanov:1981xt}
  V.~F.~Mukhanov and G.~V.~Chibisov,
  JETP Lett.\  {\bf 33} (1981) 532
   [Pisma Zh.\ Eksp.\ Teor.\ Fiz.\  {\bf 33} (1981) 549].
\bibitem{Kehagias:2014sda}
  A.~Kehagias and M.~Maggiore,
  JHEP {\bf 1408} (2014) 029
  doi:10.1007/JHEP08(2014)029
  [arXiv:1401.8289 [hep-th]].
  \bibitem{Maggiore:2014sia}
  M.~Maggiore and M.~Mancarella,
  Phys.\ Rev.\ D {\bf 90} (2014) no.2,  023005
  doi:10.1103/PhysRevD.90.023005
  [arXiv:1402.0448 [hep-th]].
  \bibitem{Conroy:2014eja}
  A.~Conroy, T.~Koivisto, A.~Mazumdar and A.~Teimouri,
  Class.\ Quant.\ Grav.\  {\bf 32} (2015) no.1,  015024
  doi:10.1088/0264-9381/32/1/015024
  \bibitem{VanNieuwenhuizen:1973fi}
  P.~Van Nieuwenhuizen,
  Nucl.\ Phys.\ B {\bf 60} (1973) 478.
  doi:10.1016/0550-3213(73)90194-6
  \bibitem{Biswas:2013kla}
  T.~Biswas, T.~Koivisto and A.~Mazumdar,
  arXiv:1302.0532 [gr-qc].
  \bibitem{Yu:2017uyd}
  S.~Yu, C.~Gao and M.~Liu,
  arXiv:1711.04064 [gr-qc].
  \bibitem{Codello:2015mba}
  A.~Codello and R.~K.~Jain,
  Class.\ Quant.\ Grav.\  {\bf 33} (2016) no.22,  225006
  doi:10.1088/0264-9381/33/22/225006
  [arXiv:1507.06308 [gr-qc]].
\end{thebibliography}
\end{document}